# A Review on Deep Learning Techniques for the Diagnosis of Novel Coronavirus (COVID-19)

Md. Milon Islam, Fakhri Karray, *Fellow, IEEE*, Reda Alhajj, *Senior Member, IEEE*, and Jia Zeng

*Abstract*—**Novel coronavirus (COVID-19) outbreak, has raised a calamitous situation all over the world and has become one of the most acute and severe ailments in the past hundred years. The prevalence rate of COVID-19 is rapidly rising every day throughout the globe. Although no vaccines for this pandemic have been discovered yet, deep learning techniques proved themselves to be a powerful tool in the arsenal used by clinicians for the automatic diagnosis of COVID-19. This paper aims to overview the recently developed systems based on deep learning techniques using different medical imaging modalities like Computer Tomography (CT) and X-ray. This review specifically discusses the systems developed for COVID-19 diagnosis using deep learning techniques and provides insights on well-known data sets used to train these networks. It also highlights the data partitioning techniques and various performance measures developed by researchers in this field. A taxonomy is drawn to categorize the recent works for proper insight. Finally, we conclude by addressing the challenges associated with the use of deep learning methods for COVID-19 detection and probable future trends in this research area. This paper is intended to provide experts (medical or otherwise) and technicians with new insights into the ways deep learning techniques are used in this regard and how they potentially further works in combatting the outbreak of COVID-19.**

## I. INTRODUCTION

NOVEL coronavirus (COVID-19), resulting from a severe acute respiratory syndrome coronavirus 2 (SARS-CoV-2), has become a pandemic worldwide in recent times [1], [2]. The number of infected cases as well as the death rate is increasing rapidly. It is reported that about 19,000,000 people have been infected with COVID-19, the death cases are around 700,000, and the number of recovered patients are around 10,000,000 globally [3]. The universal transmission of COVID-19 has put a large amount of the world's population into quarantine, and ravaged numerous industrial sectors which in turn caused a worldwide financial crisis.

The most typical signs of the novel coronavirus include fever, dry cough, myalgia, dyspnea, and headache [4], [5] but in some scenarios, no symptoms are visible (asymptomatic) that make the disease an even bigger threat to public health. The reverse transcript polymerase chain reaction (RT-PCR) is considered as the gold standard for COVID-19 diagnosis [6]. However, the lack of resources and strict test environment requirements restrict fast and effective screening of suspicious cases. Furthermore, RT-PCR inspection also experiences high false negative rates [7]. Unfortunately, the only solution to effectively combat this transmissible disease, is through clinical vaccines as well as precise drug/therapy practices, which are not yet available.

COVID-19 has proven to be amongst the most dangerous ailments that have posed severe threat to human civilization. With the evolution of modern technology in the past few decades, ingenious solutions have been created to assist disease diagnosis, prevention as well as control which leverage smart healthcare tools and facilities [8], [9], [10], [11]. Specifically, for COVID-19 diagnosis, different imaging modalities like CT and X-ray are considered among the most effective techniques [12], [13], [14], when available, CT screening is preferred in comparison with X-rays because of its versatility and three-dimensional pulmonary view [15], [16] though X-rays are must more affordable and widely available. These traditional medical imaging modalities play a vital role in the control of the pandemic.

Artificial Intelligence (AI), an evolving software technology in the area of medical image analysis has also directly helped combating the novel coronavirus [17], [18], [19] by efficiently providing high quality diagnosis results and dramatically reducing or eliminating man power. Very recently, deep learning and machine learning, two major areas of AI have become very popular in medical applications. Deep learning based support systems are developed for COVID-19 diagnosis using both CT and X-ray samples [20], [21], [22], [23]. Some of the systems are developed based on pre-trained model with transfer learning [24], [25] and a few of them are introduced using customized networks [26], [27], [28]. Machine learning [29], [30], and data science [31] are also the diverse areas that are actively used for corona diagnosis, prognosis, prediction,

Md. Milon Islam and Fakhri Karray are with the Centre for Pattern Analysis and Machine Intelligence, Department of Electrical and Computer Engineering, University of Waterloo, ON, Canada N2L 3G1 (e-mail: milonislam@uwaterloo.ca, karray@uwaterloo.ca).

Reda Alhajj is with the Department of Computer Science, University of Calgary, Calgary, Alberta, Canada (e-mail: alhajj@ucalgary.ca).

Jia Zeng is with the Institute for Personalized Cancer Therapy, MD Anderson Cancer Center, 6565 MD Anderson Blvd, Houston, TX 77030, USA (e-mail: JZeng@mdanderson.org).

and the outbreak forecasting. Computer vision [32] has also contributed for the reduction of the severity of this pandemic. Moreover, Internet of things (IoT) [33], [34], big data [35], [36], and smartphone technology [37], [38] are extensively utilized to enable innovative solutions to fight against the spread of COVID-19.

The main aim of the paper is to review the recent developments of deep learning based COVID-19 diagnosis systems based upon data collected from medical imaging samples. A taxonomy is presented that classifies the reviewed systems based on pre-trained model with deep transfer learning and customized deep learning technique. We review the most vital schemes developed for the diagnosis of COVID-19 highlighting some aspects such as the data used for experiments, the data splitting technique, and the evaluation metrics. An open discussion with the challenges of existing deep learning based systems as well as a projection of future works is also presented.

The rest of the paper is as follows. Section II categorizes the reviewed systems for proper understanding. Section III explains the recent developed systems for COVID-19 diagnosis from both CT and X-ray samples using pre-trained model with deep transfer learning. Section IV demonstrates the custom network based COVID-19 diagnosis systems from both CT and X-ray. The discussion with challenges as well as possible future trends are depicted in section V. Lastly, the paper is concluded in section VI.

## II. TAXONOMY OF DEEP LEARNING BASED COVID-19 DIAGNOSIS SYSTEMS

Deep learning techniques are able to explain complex problems by learning from simple depictions. The main features that have made the deep learning methods popular are the capability of learning the exact representations and the property of learning the data in a deep manner where multiple layers are utilized sequentially [39], [40]. Deep learning methods are widely used in medical systems such as biomedicine [41], smart healthcare [42], drug discovery [43], medical image analysis [44], etc.

More recently, it is extensively used in the automated diagnosis of COVID-19 patients. In general, deep learning based systems are comprised of several steps such as data collection, data preparation, feature extraction and classification, and performance evaluation. The general pipeline of a COVID-19 diagnosis system based on deep learning is illustrated in Fig. 1. At the data collection stage, the patients from the hospital environment are considered as a participant. The data may have different forms but for COVID-19 diagnosis, imaging techniques like CT and X-ray samples are taken. The following necessary step is the data preparation that converts the data into an appropriate format. In this step, data pre-processing includes some operations like noise removal, resizing, augmentation, and so on. The data partitioning step splits the data into training, validation, and testing set for the experiment. Generally, cross-validation technique is utilized for data partitioning. The training data is used to develop a particular model that is evaluated by validation data, and the performance of the developed model is appraised by test data. The major step of deep learning based COVID-19 diagnosis is the feature extraction and classification. In this stage, the deep learning technique automatically extracts the feature performing several operations repeatedly, and finally, the classification is done based on class labels (healthy or COVID-19). Lastly, the developed system is assessed by some evaluation metrics like accuracy, sensitivity, specificity, precision, F1-score, and so on.

In this paper, a taxonomy of classifying COVID-19 diagnosis system is presented to facilitate the navigation of the landscape. Two different perspectives are applied which are related to the used deep learning techniques and the used imaging modalities (see Fig. 2). In this paper, we have reviewed a total of 45 COVID-19 diagnosis systems. Among them, 23 systems (51.11% of the total reviewed systems) used pre-trained model for diagnosis purposes and 22 (48.89% of the total reviewed systems) used custom deep learning techniques for COVID-19 diagnosis. From a different perspective, 25 reviewed systems used X-ray images (55.55% of the total reviewed systems) as data source, and the remaining 20 systems utilized CT scans (44.44% of the total reviewed systems).

## III. PRE-TRAINED MODEL WITH DEEP TRANSFER LEARNING

A pre-trained model is one that has already been trained in fields similar to the context of the application. In transfer learning, weight and bias are transferred from a large trained model to a similar new model for testing or retraining. There are several advantages for using pre-trained models with deep transfer learning. In general, training a model from scratch for large dataset requires high computing power and is time consuming [45], [46]. The pre-trained model with transfer learning enables the facility to speed up the convergence with network generalization [47], [48]. Numerous pre-trained models that are utilized in transfer learning are designed for the large convolutional neural network (CNN). There are several pre-trained models which are used for COVID-19 diagnosis such as AlexNet [49], GoogleNet [50], SqueezeNet [51], different versions of Visual Geometry Group (VGG) [52], diverse kinds of ResNet [53], Xception [54], different forms of inception [55], diverse types of MobileNet [56], DenseNet [57], U-Net [58], etc. The systems developed for COVID-19 diagnosis are described next.

### A. Diagnosis Using Computer Tomography (CT) Images

#### 1) Diagnosis Based on Multiple Source Data

Wu et al. [59] introduced a deep learning based coronavirus screening framework using the concept of multi-view fusion. The system used a variant of CNN called ResNet50. The dataset is collected from two different hospitals in China. A total of 495 images are taken into account for the experiment in which 368 are associated with confirmed COVID-19 cases, and 127 are of other pneumonia. In this scheme, the dataset is divided into a proportion of 80%, 10%, and 10% for training, testing, and validation respectively. Each of the images, considered in the system are resized into 256×256 before the network development. In the test case, the developed system obtained accuracy of 76%, sensitivity of 81.1%, specificity of 61.5%, and Area under Curve (AUC) of 81.9%. The results are analysed both for single-view and multi-view fusion model but the multi-view fusion model demonstrates a superior performance. In another research, Xu et al. [60] developed a system for classifying healthy individuals from COVID-19 pneumonia and Influenza-A viral pneumonia utilizing CNN variants. The used pre-trained model in this system is Resnet18. The data is collected from three different hospitals in China. This study considers 618 CT images in which 219 images are obtained from patients infected with COVID-19, 224 from Influenza-A viral pneumonia, and 175 from normal individuals. To train the model, a total of 85.4% (528) images are used, and the remaining samples are used to test the developed model. The framework achieved 86.7% accuracy, 81.5% sensitivity, 80.8% precision, and 81.1% F1-score from the experiment.

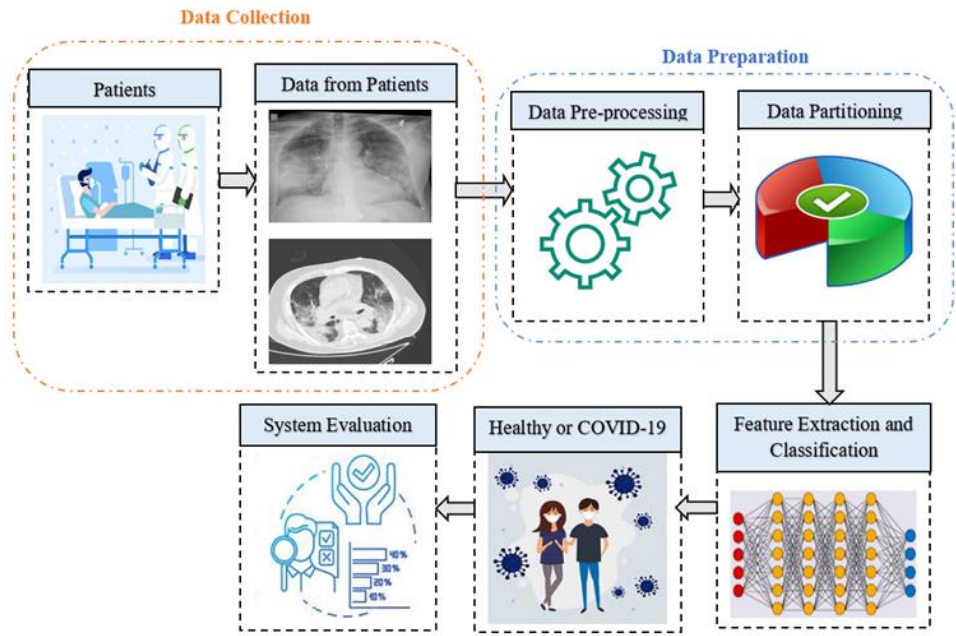

Fig. 1. The general pipeline of deep learning based COVID-19 diagnosis system.

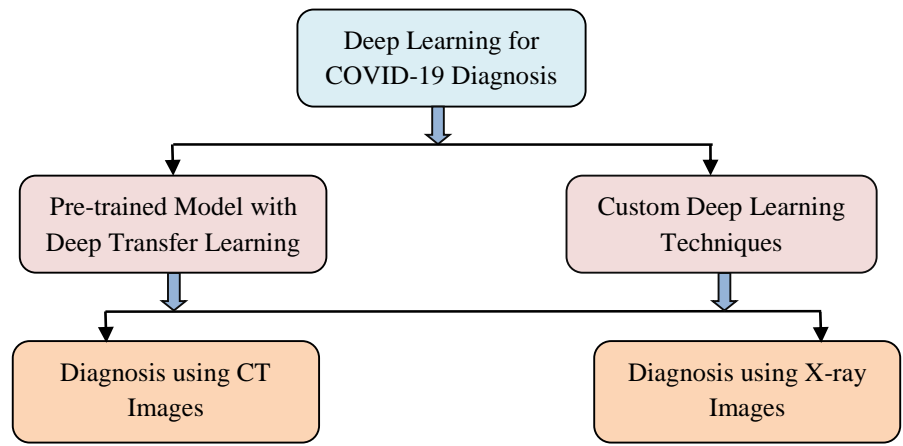

Fig. 2. Taxonomy of the recent developed COVID-19 diagnosis systems using deep learning.

Afterward, Jin et al. [61] developed an artificial intelligence based coronavirus diagnosis system using a variant of CNN named ResNet152. The pre-trained model used 152 convolutional, subsampling, and fully-connected layers. The used dataset is collected from three renowned hospitals of China, and two publicly available databases. A total number of 1881 of cases are considered where 496 cases are for COVID-19 infected patients, and 1385 are negative cases. The dataset is split randomly for experiments. The system achieved an accuracy of 94.98%, sensitivity of 94.06%, specificity of 95.47%, precision of 91.53%, F1-score of 92.78, and AUC of 97.91% from the experiment. Furthermore, Jin et al. [62] introduced a medical system for COVID-19 screening using deep learning techniques. Their system used various pre-trained models of CNN like DPN-92, Inception-v3, ResNet-50, and Attention ResNet-50 with 3D U-Net++. The dataset is retrieved from different five hospitals in China. In this system, a total of 139 samples are used where 850 samples from COVID-19, and 541 samples from other cases which are considered as negative. The total data is randomly split into training and testing sets for performance evaluation. As the evaluation metrics, the system obtained sensitivity, specificity, and AUC of 97.4%, 92.2%, and 99.1% respectively using 3D Unet++-ResNet-50 which is considered as the best model. In another research work, Li et al. [63] demonstrated an automatic system (COVNet) for the diagnosis of coronavirus from CT images using deep learning technique which is a variant of CNN named ResNet50. The used dataset consists of 4536 chest CT samples, including 1296 samples for COVID-19, 1735 for community-acquired pneumonia (CAP), and 1325 for non-pneumonia. The dataset is partitioned into training and testing set in a proportion of 90% and 10% respectively. The experimental result showed that the system obtained sensitivity of 90%, specificity of 96%, and AUC of 96% for COVID-19 cases.

Moreover, Javaheri et al. [64] developed a deep learning approach called CovidCTNet for detecting coronavirus infection via CT images. The system used BCDU-Net architecture which is developed based on U-Net. The scheme distinguished COVID-19 from CAP as well as other lung disorders. For the experiment, the system used 89,145 CT images in total where 32,230 CT slices are confirmed with COVID-19, 25,699 CT slices are confirmed with CAP, and 31,216 CT slices are with healthy lungs or other disorder. The dataset is partitioned using hold-out method i.e. 90% is used for training and 10% is utilized for testing. It is obvious from the experimental results that the developed system obtained accuracy, sensitivity, specificity, AUC of 91.66%, 87.5%, 94%, and 95% respectively. For the proper diagnosis of COVID-19, Yousefzadeh et al. [65] introduced a deep learning framework called ai-corona which is worked based on CT images. The system is comprised of several variants of CNN named DenseNet, ResNet, Xception, and EfcientNetB0. The used dataset contained 2124 CT slices in overall where 1418 images are of non-COVID-19, and 706 slices are of COVID-19 infected cases. The dataset maintained a ratio of 80% and 20% for training and validation set respectively. The proposed system found accuracy of 96.4%, sensitivity of 92.4%, specificity of 98.3%, F1-score of 95.3%, and AUC of 98.9% from the experiment.

### 2) Diagnosis Based on Single Source Data

Ardakani et al. [66] proposed a system for the detection of COVID-19 using ten variants of CNN techniques in CT images. The used popular variants for diagnosis are AlexNet, VGG-16, VGG-19, SqueezeNet, GoogleNet, MobileNet-V2, ResNet-18, ResNet-50, ResNet-101, and Xception. In the proposed system, a total of 1020 CT samples are considered from the cases of COVID-19 and non-COVID-19. The dataset is split into training and validation set in a proportion of 80% and 20% respectively. Among the 10 networks, ResNet-101 and Xception performed comparatively better than the others. It is evident from the experimental results that the ResNet-101 model obtained accuracy of 99.51%, sensitivity of 100%, AUC of 99.4%, and specificity of 99.02%. In other network, Xception found the accuracy, sensitivity, AUC, and specificity of 99.02%, 98.04%, 87.3%, and 100% respectively. In another study, Chen et al. [67] introduced a deep learning based scheme for the COVID-19 detection of high-resolution CT images where they used a powerful pre-trained model named UNet++ for detection. Initially, UNet++ extracted valid region in CT images. In this study, 46,096 images are collected from a hospital including 51 COVID-19 infected patients and 55 infected with other diseases. Among the dataset, 35,355 images are selected while eliminating low images using filtering, and partitioned into training and testing set respectively. Sensitivity of 94.34%, specificity of 99.16%, accuracy of 98.85%, precision of 88.37%, and negative predictive value (NPV) of 99.61% are achieved. Further, Cifci [68] presented a scheme for the early diagnosis of coronavirus using pre-trained models with deep transfer learning. The pre-trained models are AlexNet and Inception-V4 which are popular for medical image analysis. The study is carried out through CT images. To develop the system, 5800 CT images are retrieved from a public repository. As a training step, 4640 (80%) CT samples are used, while 1160 (20%) samples are used for testing. AlexNet performed comparatively better than Inception-V4 which is found through experimental results. AlexNet got an overall accuracy of 94.74% with sensitivity, and specificity of 87.37%, and 87.45% respectively.

Table I summarizes the aforementioned deep learning based COVID-19 diagnosis systems from CT samples using pre-trained model with deep transfer learning and describes some of the significant factors, such as data sources, number of images and classes, data partitioning technique, the used techniques for diagnosis, and the performance measures of the developed systems.

### B. Diagnosis Using X-ray Images

#### 1) Diagnosis Based on Multiple Source Data

Apostolopoulos and Bessiana [69] developed a system for the automatic diagnosis of COVID-19 cases utilizing the concept of transfer learning with five variants of CNNs. The pre-trained models which are used in the system are VGG19, MobileNetv2, Inception, Xception, and Inception-ResNetv2. The system considered 1427 images including 224 for COVID-19, 700 for common pneumonia, and 504 for healthy cases in the first scenario. In the second scenario, 224 COVID-19 images, 714 bacterial and viral pneumonia images, and 504 healthy individual images are considered. The dataset was divided using the 10-fold cross-validation method. It was revealed that the highest accuracy of 96.78%, sensitivity of 98.66%, and specificity of 96.46% are obtained for the second dataset using MobileNetv2. In another research work, Loey et al. [70] introduced a novel system for the diagnosis of coronavirus using Generative Adversarial Network (GAN) and pre-trained models of CNN with deep transfer learning. The pre-trained models which are used in the proposed system are Alexnet, Googlenet, and Resnet18. As the number of X-ray images for COVID-19 is small, GAN is used to generate more samples for accurate detection of this virus. A total number of 307 images are considered including four classes like COVID-19, normal, pneumonia_bac, and pneumonia_vir.

TABLE I
SUMMARY OF DEEP LEARNING BASED COVID-19 DIAGNOSIS IN CT IMAGES USING PRE-TRAINED MODEL WITH DEEP TRANSFER LEARNING

| Authors | Data Sources | No. of images | No. of classes | Partitioning | Techniques | Performances (%) |
|---------|-------------|---------------|----------------|--------------|------------|-------------------|
| Wu et al. [59] | Two different hospitals (China Medical University, Beijing Youan Hospital) | 495 (COVID-19=368, other pneumonia=127) | 2 (COVID-19, other pneumonia) | Training=80%, Validation=10%, Testing=10% | ResNet50 | Accuracy=76, Sensitivity=81.1, Specificity=61.5, AUC=81.9 |
| Xu et al. [60] | Zhejiang University, Hospital of Wenzhou, Hospital of Wenling | 618 (COVID-19=219, Influenza-A-viral-pneumonia=224, irrelevant-to-infection=175) | 3 (COVID-19, Influenza-A-viral-pneumonia, irrelevant-to-infection) | Training + Validation=85.4%, Testing=14.6% | ResNet18 | Accuracy=86.7, Sensitivity=81.5, Precision=80.8, F1-Score=81.1 |
| Jin et al. [61] | Three different hospitals (Wuhan Union Hospital, Western Campus of Wuhan Union Hospital, Jianghan Mobile Cabin Hospital), LIDC-IDRI [71], ILD-HUG [72] | 1881 (COVID-19 positive=496, COVID-19 negative=1385) | 2 (COVID-19 positive, COVID-19 negative) | Random partition | ResNet152 | Accuracy=94.98, Sensitivity=94.06, Specificity=95.47, Precision=91.53, F1-Score=92.78, AUC=97.91, NPV=96.86, Youden Index=89.53 |
| Jin et al. [62] | Five different hospitals of China | 1391 (COVID-19 positive=850, COVID-19 negative=541) | 2 (COVID-19 positive, COVID-19 negative) | Random partition | DPN-92, Inception-v3, ResNet-50, Attention ResNet-50 with 3D U-Net++ | Sensitivity=97.04, Specificity=92.2, AUC=99.1 |
| Li et al. [63] | Multiple hospitals environment | 4536 (COVID-19=1296, CAP=1735, non-pneumonia=1325) | 3 (COVID-19, CAP, non-pneumonia) | Training=90%, Testing=10% | ResNet50 | Sensitivity= 90, Specificity=96, AUC=96 |
| Javaheri et al. [64] | Five medical centers in Iran, SPIE-AAPM-NCI [73], LUNGx [74] | 89,145 (COVID-19=32,230, CAP=25,699, other diseases=31,216) | 3 (COVID-19, CAP, other diseases) | Training=90%, Validation=10% | BCDU-Net (U-Net) | Accuracy=91.66, Sensitivity=87.5, Specificity=94, AUC=95 |
| Yousefzadeh et al. [65] | Real-time data from hospital environment | 2124 (COVID-19=706, non-COVID-19=1418) | 2 (COVID-19, non-COVID-19) | Training=80%, Validation=20% | DenseNet, ResNet, Xception, EcientNetB0 | Accuracy=96.4, Sensitivity= 92.4, Specificity=98.3, F1-Score= 95.3, AUC=98.9, Kappa=91.7 |
| Ardakani et al. [66] | Real-time data from hospital environment | 1020 (COVID-19=510, non-COVID-19=510) | 2 (COVID-19, non-COVID-19) | Training=80%, Validation=20% | AlexNet, VGG-16, VGG-19, SqueezeNet, GoogleNet, MobileNet-V2, ResNet-18, ResNet-50, ResNet-101, Xception | Accuracy=99.51, Sensitivity=100, Specificity=99.02, Precision=99.27, AUC=99.4, NPV=100 |

| Chen et al. [67] | Renmin Hospital of Wuhan University | 35,355 | 2 (COVID-19, other diseases) | Random partition | UNet++ | Accuracy=98.85, Sensitivity=94.34, Specificity=99.16, Precision=88.37, AUC=99.4, NPV=99.61 |
|---|---|---|---|---|---|---|
| Cifci [68] | kaggle.com (benchmark web of dataset science) | 5800 | 2 (COVID-19, other pneumonia) | Training=80%, Testing=20% | AlexNet, Inception-V4 | Accuracy=94.74, Sensitivity=87.37, Specificity=87.45 |

The system experimented on three different scenarios of dataset depending on the consideration of class level. Considering four classes, Googlenet obtained the highest accuracy of 80.6%. Alexnet and Googlenet achieved accuracy of 85.2% and 100% respectively considering three and two classes.

Horry et al. [75] described a COVID-19 detection framework using the concept of pre-trained model in X-ray images. The proposed system used four popular pre-trained models like VGG, Inception, Xception, and Resnet with transfer learning. The used dataset consisted of 100 COVID-19 cases, 100 pneumonia, and 200 healthy cases for experiments. In this system, a ratio of 80:20 is preserved for training and testing set as a data partition. The experimental findings reveal that the system obtained precision, sensitivity, and F1-score of 83%, 80%, and 80% respectively using VGG-19 which is measured as the highest performance in the study considering three-class data. Further, Ozcan [76] proposed a deep learning scheme with a combination of the grid search strategy and three pre-trained models of CNN. The used pre-trained models are GoogleNet, ResNet18, and ResNet50. The grid search technique is used to select the best hyperparameter and the pre-trained models are utilized for feature extraction and classification. The system used three public datasets where the images are of 242 bacteria cases, 131 COVID-19 cases, 200 normal cases, and 148 viral cases. All the data are partitioned into training, testing, and validation set in a proportion of 50:30:20. The ResNet50 with grid search performed better and obtained accuracy of 97.69%, sensitivity of 97.26%, specificity of 97.90%, precision of 95.95%, and F1-score of 96.60%.

Sethy and Behra [77] introduced a system for the diagnosis of COVID-19 cases using pre-trained models of CNN and Support Vector Machine (SVM). The algorithm used eleven CNN pre-trained models for automatic extraction of features, and SVM for classification. In this system, two separate datasets were used where the first dataset included 25 positive COVID-19 and 25 negative X-ray images of COVID-19. A total of 133 images containing Middle East Respiratory Syndrome (MERS), SARS, and Acute Respiratory Distress Syndrome (ARDS) are used as positive samples and 133 normal X-ray images as negative samples in the second dataset. From the experimental results, it is found that Resnet50 with SVM obtained accuracy, False Positive Rate (FPR), Matthews Correlation Coefficient (MCC), and Kappa of 95.38%, 95.52%, 91.41%, and 90.76% respectively which the best is in the developed system for the first scenario of the dataset. Minaee et al. [78] proposed a framework named Deep-COVID using the concept of deep transfer learning for COVID-19 prediction in X-ray images. Four popular pre-trained models like ResNet18, ResNet50, SqueezeNet, and DenseNet-121 were considered in this study for COVID-19 diagnosis. In total, 5071 images are collected from different open-access resources. Among them, 2000 images with 31 COVID-19 cases were used for training, and 3000 images with 40 COVID-19 infected cases were used for testing in the experiments. The resulting dataset was named COVID-Xray-5k. The best performance obtained by the system is sensitivity of 100%, and specificity of 95.6% using SqueezeNet.

In another study, Punn and Agarwal [79] developed an automated COVID-19 diagnosis system using several pre-trained models like ResNet, Inception-v3, Inception ResNet-v2, DenseNet169, and NASNetLarge with a small number of X-ray images. The system used random oversampling and weighted class loss function for fine-tuning called transfer learning. In this system, a total of 1076 chest X-ray images are considered for experiments. The dataset is partitioned into 80%, 10%, and 10% ratio for training, testing, and validation set respectively. From the experimental results, it was shown that NASNetLarge performed comparatively better and achieved accuracy, precision, sensitivity, AUC, specificity, and F1-score of 98%, 88%, 91%, 99%, 98%, and 89% respectively. Afterward, Narin et al. [80] introduced a method for automatically classifying COVID-19 infected patients from X-ray images using the variants of CNN. The pre-trained models used are ResNet50, InceptionV3, and Inception-ResNetV2 which obtained higher predictive accuracy on a subset X-ray dataset. The system used a total of 100 X-ray images where 50 images were from COVID-19 patients while the remaining 50 from healthy individuals. The 5-fold cross-validation was used to partition the dataset for the experiment. The system achieved an accuracy of 98%, 97%, and 87% from ResNet50, InceptionV3, and Inception-ResNetV2 respectively in test cases. In terms of other evaluation metrics, the best performance was obtained using RecNet50 with a recall of 96%, specificity of 100%, precision of 100%, and F1-score of 98%.

Bukharia et al. [81] presented a COVID-19 diagnosis system using a variant of CNN named Resnet50. The system considered 278 X-ray images of three classes where 89 samples of COVID-19 infected, 93 samples of healthy participants, and 96 samples of pneumonia patients. The collected dataset was split into two sets like training and testing in a proportion of 80% (223 images), and 20% (55 images). The diagnosis process obtained accuracy, precision, recall, and F1-score of 98.18 %, 98.14%, 98.24%, and 98.19 % respectively from the experiment.

TABLE II
Summary of Deep Learning Based COVID-19 Diagnosis in X-ray Images Using Pre-Trained Model with Deep Transfer Learning

| Authors | Data Sources | No. of images | No. of classes | Partitioning | Techniques | Performances (%) |
|---|---|---|---|---|---|---|
| Apostolopoulos and Bessiana [69] | COVID-19 X-ray image database [82], Kaggle dataset [83], Kermany et al. [84] | 1442 (COVID-19=224, pneumonia=714, normal=504) | 3 (COVID-19, pneumonia, normal) | 10- fold cross-validation | VGG19, MobileNetv2, Inception, Xception, Inception-ResNetv2 | Accuracy=96.78, Sensitivity=98.66, Specificity=96.46 |
| Loey et al. [70] | COVID-19 X-ray image database [82], Kermany et al. [84], Dataset [85] | 307 (COVID=69, normal=79, pneumonia_bac=79, pneumonia_vir=79) | 4 (COVID, normal, pneumonia_ba, pneumonia_vir) | Training=80%, Testing=10%, Validation=10% | GAN, Alexnet, Googlenet, Resnet18 | Accuracy=100, Sensitivity= 100, Precision= 100, F1-Score= 100 |
| Horry et al. [75] | COVID-19 X-ray image database [82], NIH Chest X-Ray [86] | 400 (COVID-19=100, pneumonia=100, normal=200) | 3 (COVID-19, pneumonia, normal) | Training=80%, Testing=20% | VGG16, VGG19, ResNet50, InceptionV3, Xception | Sensitivity=80, Precision=83, F1-Score=80 |
| Ozcan [76] | COVID-19 X-ray image database [82], Kaggle chest x-ray repository [87], Italian Society of Medical and Interventional Radiology : COVID-19 Database [88] | 721 (COVID-19=131, bacteria=242, normal=200, virus=148) | 4 (COVID-19, normal, bacteria, virus) | Training=50%, Testing=30%, Validation= 20% | GoogleNet, ResNet18, ResNet50 | Accuracy=97.69, Sensitivity= 97.26, Specificity= 97.90, Precision= 95.95, F1-Score= 96.60 |
| Sethy and Behra. [77] | COVID-19 X-ray image database [82], NIH Chest X-Ray [86], Kaggle chest x-ray repository [87] | 316 | 2(COVID-19+, COVID-19-) | Training=60%, Testing=20%, Validation= 20% | AlexNet, VGG16, VGG19, GoogleNet, ResNet18, ResNet50, ResNet101, InceptionV3, InceptionResNet V2, DenseNet201, XceptionNet, SVM | Accuracy=95.38, Sensitivity= 97.47, Specificity= 93.47, Precision= 95.95, F1-Score= 95.52, MCC=91.41, FPR=95.52, Kappa=90.76 |
| Minaee et al. [78] | COVID-19 X-ray image database [82], ChexPert [89] | 5071 ( COVID-19=71, non-COVID=5000) | 2 (COVID-19, non-COVID) | Training=40%, Testing=60% | ResNet18, ResNet50, SqueezeNet, DenseNet-121 | Sensitivity= 100, Specificity= 95.6, AUC=99.6 |
| Punn and Agarwal [79] | COVID-19 X-ray image database [82], RSNA Pneumonia Detection Challenge dataset [90] | 1076 (COVID-19=108, pneumonia=515, normal=453) | 3 (COVID-19, pneumonia, normal) | Training=80%, Testing=10%, Validation= 10% | ResNet, Inception-v3, Inception, ResNet-v2, DenseNet169, NASNetL | Accuracy=98, Sensitivity= 91, Specificity= 91, Precision= 98, F1-Score= 89, AUC=99 |
| Narin et al. [80] | COVID-19 X-ray image database [82], Kaggle chest x-ray repository [87] | 100 (COVID-19=50, normal=50) | 2 (COVID-19, normal) | 5- fold cross-validation | ResNet50, InceptionV3, Inception-ResNetV2 | Accuracy=98, Sensitivity= 96, Specificity= 100, Precision= 100, F1-Score= 98, AUC=100 |
| Bukharia et al. [81] | COVID-19 X-ray image database [82], NIH Chest X-Ray [86] | 278 (COVID-19=89, normal=93, pneumonia=96,) | 3 (COVID-19, normal, pneumonia,) | Training=80%, Testing=20% | ResNet50 | Accuracy=98.18, Sensitivity= 98.24, Precision= 98.14, F1-Score= 98.19 |

| Abbas et al. [91] | COVID-19 X-ray image database [82], Japanese Society of Radiological Technology (JSRT) [92], [93] | 196 (COVID-19=105, normal=80, SARS=11) | 3 (COVID-19, normal, SARS) | Training=70%, Testing=30% | DeTraC-ResNet18 | Accuracy=95.12, Sensitivity=97.91, Specificity=91.87, Precision=93.36 |
|---|---|---|---|---|---|---|
| Moutounet-Cartan [94] | COVID-19 X-ray image database [82], Kermany et al. [84] | 327 (COVID-19=125, normal=152, pneumonia=50) | 3 (COVID-19, normal, pneumonia) | 5-fold cross-validation | VGG16, VGG19, InceptionResNet V2, InceptionV3, Xception | Accuracy= 84.1, Sensitivity=87.7, AUC=97.4 |
| Maguolo and Nanni [95] | COVID-19 X-ray image database [82], Kaggle chest x-ray repository [87], ChexPert [89], ChestX-ray8 [96] | 339,271 (COVID-19=144, pneumonia=339, 127) | 2 (COVID-19, pneumonia) | 10-fold cross-validation | AlexNet | AUC=99.97 |
| Hemdan et al. [97] | COVID-19 X-ray image database [82] | 50 (COVID-19 =25, normal = 25) | 2 (COVID-19, normal) | Training=80%, Testing=20% | VGG19, DenseNet121, InceptionV3, ResNetV2, Inception-ResNet-V2, Xception, MobileNetV2 | Accuracy=90, Sensitivity= 100, Specificity= 100, Precision= 100, F1-Score= 91, AUC=90 |

Moreover, Abbas et al. [91] categorized COVID-19 infected patients, from healthy individuals using Decompose, Transfer, and Compose (DeTraC) deep ResNet18. The proposed DeTraC can fix any anomalies in the image dataset by using a class decomposition method to investigate its class boundaries. In this system, a total of 196 images were utilized where 80 samples of normal patients, 105 samples of COVID-19, and 11 samples of SARS. The system generated 1764 samples from given samples using decomposition. The dataset was split into two groups, 70% for system training and 30% for evaluation. The proposed system achieved accuracy of 95.12%, sensitivity of 97.91%, specificity of 91.87%, and precision of 93.36% using DeTraC-ResNet18 framework.

In another research project, Moutounet-Cartan [94] developed a deep learning based system to diagnose the novel coronavirus as well as other pneumonia diseases from X-ray images. The system used the following variants of CNN architecture named VGG-16, VGG-19, InceptionResNetV2, InceptionV3, and Xception for diagnosis. In this study, in total 327 X-ray images were taken where 152 cases were from healthy people, 125 from COVID-19 cases, and the remaining 50 cases from other pneumonia diseases. The dataset is partitioned using the principle of 5-fold cross-validation. The system found VGG-16 as the best performing model and obtained overall accuracy of 84.1%, sensitivity 87.7%, and AUC of 97.4% where the sensitivity and AUC were considered only for COVID-19 cases. Furthermore, Maguolo and Nanni [95] evaluated the performance of COVID-19 detection system from X-ray samples utilizing a popular pertained model named AlexNet. The system used four different publicly available datasets to evaluate the performance. A total of 339,271 images were taken where 144 images for COVID-19 patients, 108,948 samples of pneumonia and bacteria except COVID-19, 224,316 chest radiographs of bacteria and pneumonia, and 5,863 paediatric images viral and bacterial pneumonia. The dataset was partitioned into 10-fold cross-validation for training and testing. Using the concept of deep transfer learning, the system obtained the highest AUC of 99.97% in the study.

### 2) Diagnosis Based on Single Source Data

Very recently, Hemdan et al. [97] proposed a system named COVIDX-Net to diagnose coronavirus using the variants of CNN in X-ray images. A total of seven pre-trained models are considered in this study. The dataset consisted of 50 images where 25 images are from healthy people and the remaining 25 samples from COVID-19 cases. For the experiment, the dataset was split into a proportion of 80% and 20% for training and testing set respectively. The experimental results revealed that VGG-19 and DenseNet outperformed the other pre-trained models with an accuracy of 90% and F1-score of 91%. InceptionV3 obtained the worst results.

Table II summarizes the aforementioned deep learning based COVID-19 diagnosis systems from X-ray samples using pre-trained model with deep transfer learning and describes some of the significant factors, such as data sources, number of images and classes, data partitioning technique, the used techniques for diagnosis, and the performance measures of the reported systems.

## IV. CUSTOM DEEP LEARNING TECHNIQUES

Custom deep learning techniques provide the ability to develop a user-friendly architecture and to allow for more consistent and accurate performance due to the attention to the specific application of interest. The custom networks are evolved with the use of a particular deep learning method [98] or the hybridization of deep learning algorithms [99], [100] or the hybridization of deep learning with other fields of AI such as machine learning, data mining, nature-inspired algorithms, etc. [101], [102]. No previous weights and bias are used in the customized network like pre-trained model hence it requires comparatively high computation power and time. The systems developed for COVID-19 diagnosis are outlined as follows.

### A. Diagnosis Using Computer Tomography (CT) Images

#### 1) Diagnosis Based on Multiple Source Data

Elghamrawy and Hassanien [103] proposed a scheme for the diagnosis and prediction of coronavirus infected patients using a combination of CNN and Whale Optimization Algorithm (WOA) from CT samples. In the proposed method, CNN is used for diagnosis, and WOA is utilized for prediction. The used dataset is collected from publicly available databases consisted of 617 CT scans. Among them, 134 images are excluded as it contains non-lung region. A total of 432 images confirmed of COVID-19 and 151 cases of other viral pneumonia were considered. To achieve better performance, the dataset was divided into a proportion of 65%, and 35% for training and testing respectively. The proposed system obtained overall accuracy, sensitivity, and precision of 96.40%, 97.25%, and 97.3% respectively for diagnosis. Further, He et al. [104] proposed a deep learning method named CRNet for the detection of COVID-19 using CT images. In this system, a total of 746 CT images were analysed where 349 were associated with COVID-19 cases, and 397 with non-COVID-19 cases. The dataset was formed by merging three publicly available datasets which was divided into three sets named training, testing, and validation set in a proportion of 60%, 25%, and 15% respectively. The proposed system obtained accuracy of 86%, F1-score of 85%, and AUC of 94% from the experimental results. In comparison with other prominent pre-trained models, the proposed system used comparatively less tuning parameters.

Afterwards, Wang et al. [105] introduced a scheme for COVID-19 diagnosis using a modified CNN technique named modified-Inception. The basic difference between Inception and modified-Inception is that modified-Inception reduces the dimension of attributes before final classification. In the experiment, the scheme used 1040 CT images, in which 740 were tagged as COVID-19 positive and 325 as COVID-19 negative. The dataset was partitioned into training, testing, and validation set randomly. The experimental outcomes revealed that the scheme achieved accuracy, sensitivity, specificity, precision, and F1-score of 79.3%, 83%, 67%, 55%, and 63% respectively on the testing samples. Moreover, Liu et al. [106] developed an automatic COVID-19 diagnosis system using deep learning method via CT images. The system used modified DenseNet-264 (COVIDNet) for diagnosis where the model consisted of 4 dense blocks. In this system, 920 COVID-19 and 1,073 non-COVID-19 cases were considered for the experiment. To obtain better performance, the dataset is partitioned into three sets namely training, testing, and validation in a proportion of 60%, 20%, and 20% respectively. The developed system obtained accuracy of 94.3%, AUC of 98.6%, sensitivity of 93.1%, specificity of 95.1%, precision of 93.9%, NPV of 94.5%, and F1-score of 93.5%.

In another study , Ying et al. [107] introduced a deep learning technique based on the Details Relation Extraction neural network (DRE-Net) named Deep Pneumonia for the diagnosis of COVID-19 cases utilizing CT images. The dataset was collected from two popular hospitals in China. In this system, a total of 1990 image slices were taken where 777 images for

COVID-19, 505 slices for bacterial pneumonia, and 708 samples from normal people. The dataset was split in a proportion of 60%, 30%, and 10% for training, testing, and validation set respectively. The proposed system obtained accuracy of 94%, sensitivity of 93%, precision of 96%, F1-score of 94%, and AUC of 99%. To detect COVID-19, Zheng et al. [108] proposed a 3D deep convolution neural network (DeCoVNet) from CT scans. The proposed network is comprised of three segments like a vanilla 3D convolution, a batch norm layer, and a subsampling layer. The data for the study was collected from the hospital environment. A total of 630 CT samples were used for experiment where 80% (499 images) were in the training set, and the rest 20% (131images) were used in testing. From the experimental outcome, accuracy, sensitivity, specificity, precision, NPV, AUC of 90.1%, 90.7%, 91.1%, 84%, 98.2%, and 95.9% are achieved.

Hasan et al. [109] proposed a hybrid system using the concept of Q-deformed entropy and deep learning features (QDE–DF) to differentiate COVID-19 infected people from pneumonia cases, and healthy people utilizing CT images. For deep features extraction, CNN and Q-deformed entropy were used, and LSTM was used to classify the cases from deep features. A total of 321 chest CT samples were used for this study, consisting of 118 CT samples of COVID-19 cases, 96 CT samples of pneumonia cases, and 107 CT samples of healthy individuals. Approximately, 16 attributes were extracted from each image using a feature extraction technique. To assess the developed system, the dataset was partitioned in a proportion of 70%, and 30% for training and testing set respectively. The system obtained accuracy of 99.68% which is considered as the highest in this study. Further, Amyar et al. [110] developed a scheme using deep learning method to diagnose COVID-19 patients from CT samples. The system consists of an encoder for reconstruction and two decoders for segmentation, and for classification purposes, a multi-layer perceptron is used. The used dataset included 1044 cases where 449 cases were of confirmed COVID-19, 100 cases from healthy individuals, 98 samples were from confirmed lung cancer patients, and 397 from various other kinds of pathology. Collectively, 449 were associated with COVID-19 and 595 were not. The dataset was partitioned into training, validation, and testing set in a ratio of 80%, 10%, and 10% respectively. The proposed system received accuracy of 86%, sensitivity of 94%, specificity of 79%, AUC of 93%.

#### 2) Diagnosis Based on Single Source Data

Singh et al. [111] classified COVID-19 infected (positive) cases from other (negative) cases using deep learning technique (CNN). In this network, CNN's initial parameters were tuned with the application of multi-objective differential evolution (MODE). A total of 150 CT samples were taken where 75 samples for COVID-19 positive and 75 images for COVID-19 negative. Different variations in training and testing dataset ratio of 20:80 %, 30:70%, 40:60%, 50:50%, 60:40%, 70:30%, 80:20%, and 90:10%, respectively are taken to conduct the experiment. The best performed ratio for the proposed system is 90%, and 10% for training and testing set individually in the maximum cases. The system obtained accuracy of 93.25%, sensitivity of 90.70%, specificity of 90.72%, F1-score of 89.96%, and Kappa of 90.60% from the experiment. In another work, Farid et al. [119] introduced a new approach for classifying COVID-19 infection using the attributes from CT images. The image parameters were taken using four image filters in mixture with developed hybrid composite extraction method. The system considered two classes of data named COVID-19, and SARS, each of the class comprised of 51 images. The dataset was partitioned using 10-fold cross-validation technique to obtain a better outcome. The developed system obtained accuracy, precision, f1-score, and AUC of 94.11%, 99.4%, 94%, and 99.4% respectively.

TABLE III
SUMMARY OF DEEP LEARNING BASED COVID-19 DIAGNOSIS IN CT IMAGES USING CUSTOMIZED NETWORK

| Authors | Data Sources | No. of images | No. of classes | Partitioning | Techniques | Performances (%) |
|---|---|---|---|---|---|---|
| Elghamrawy and Hassanien [103] | Italian Society of Medical and Interventional Radiology : COVID-19 Database [88], COVID-CT [112] | 583 (COVID-19=432, viral pneumonias=151) | 2 (COVID-19, viral pneumonia) | Training=65%, Testing=35% | WOA-CNN | Accuracy=96.40, Sensitivity=97.25, Precision =97.3 |
| He et al. [104] | Italian Society of Medical and Interventional Radiology : COVID-19 Database [88], Covid-19 [113], Eurorad [114], Coronacases [115] | 746 (COVID-19=349, non-COVID-19⁻=397) | 2 (COVID-19, non-COVID-19) | Training=60%, Validation=15%, Testing=25% | CRNet | Accuracy=86, F1-Score=85, AUC=94 |
| Wang et al. [105] | Three different hospitals (Xi'an Jiaotong University, Nanchang University, Xi'an Medical College) | 1065 (COVID-19⁺=740, COVID-19⁻=325) | 2 (COVID-19⁺, COVID-19⁻) | Random partition | Modified-Inception | Accuracy=79.3, Sensitivity=83, Specificity=67, Precision=55, NPV=90, F1-Score=63, AUC=81, Kappa=48, Yoden index=50 |
| Liu et al. [106] | Ten designated COVID-19 hospitals in China | 1993 (COVID-19=920, non-COVID-19=1073) | 2 (COVID-19, non-COVID-19) | Training=60%, Validation=20%, Testing=20% | Modified DenseNet-264 | Accuracy=94.3, Sensitivity= 93.1, Specificity=95.1, Precision= 93.9, F1-Score= 93.5, AUC=98.6, NPV=94.5 |
| Ying et al. [107] | Two different hospitals (Hospital of Wuhan University, Third Affiliated Hospital) | 1990 (COVID-19=777, bacterial pneumonia=505, normal=708) | 3 (COVID-19, bacterial pneumonia, normal) | Training=60%, Validation=10%, Testing=30% | DRE-Net | Accuracy=94.3, Sensitivity=93, Precision=96, F1-Score=94, AUC=99 |
| Zheng et al. [108] | Three different hospitals (Union Hospital, Tongji Medical College, Huazhong University of Science and Technology) | 630 | 2 (COVID-positive, COVID-negative) | Training=80%, Testing=20% | DeCoVNet | Accuracy=90.1, Sensitivity=90.7, Specificity=91.1, Precision=84, NPV=98.2, AUC=95.9 |
| Hasan et al. [109] | COVID-19 [113], SPIE-AAPM-NCI Lung Nodule Classification Challenge Dataset [116] | 321 (COVID-19=118, pneumonia=96, healthy=107) | 3 (COVID-19, pneumonia, healthy) | Training=70%, Testing=30% | QDE–DF | Accuracy=99.68 |
| Amyar et al. [110] | COVID-CT [112], COVID-19 CT segmentation dataset [117], a hospital named Henri Becquerel Center | 1044 (COVID-19=449, non-COVID-19=595) | 2 (COVID-19, non-COVID-19) | Training=80%, Validation=10%, Testing=10% | Encoder-Decoder with multi-layer perceptron | Accuracy=86, Sensitivity=94, Specificity=79, AUC=93, |
| Singh et al. [111] | COVID-19 patient chest CT images [118] | 150 (COVID-19⁺=75, COVID-19⁻=75) | 2 (COVID-19⁺, COVID-19⁻) | Various proportions of training and testing dataset | MODE-CNN | Accuracy=93.25, Sensitivity=90.70, Specificity=90.72, F1-Score=89.96, Kappa=90.60 |
| Farid et al. [119] | Kaggle benchmark dataset [120] | 102 (COVID-19=51, SARS=51) | 2 (COVID-19, SARS) | 10-fold cross-validation | CNN | Accuracy=94.11, Precision=99.4, F1-Score=94, AUC=99.4 |

Table III summarizes the aforementioned deep learning based COVID-19 diagnosis systems from CT samples using custom deep learning techniques and demonstrates some of the important factors, such as data sources, number of images and classes, data partitioning technique, diagnosis techniques, and the evaluation metrics of the developed systems.

### B. Diagnosis Using X-ray Images

#### 1) Diagnosis Based on Multiple Source Data

Ozturk et al. [121] presented a customized network (DarkCovidNet) for the automatic diagnosis of COVID-19 in raw chest X-ray samples utilizing deep neural networks. The proposed system used DarkNet as a classifier with 17 convolutional layers. In this system, two sources of the dataset were used which includes 127 images from first sources, and 500 normal and 500 pneumonia cases from frontal X-ray samples from the second source. The dataset was partitioned in 5-fold cross-validation technique. The obtained sensitivity, specificity, precision, F1-score, and accuracy of 95.13%, 95.3%, 98.03%, 96.51%, and 98.08% respectively for binary-class which are the highest in this study. Moreover, Ucar and Korkmaz [122] developed a COVIDiagnosis-Net based on the Bayes-SqueezeNet for the diagnosis of coronavirus utilizing X-ray samples. The system used 1591 pneumonia cases with non-COVID-19, 45 COVID-19 cases, and 1203 uninfected normal patients in total as the dataset. The dataset is formed with the combination of three publicly available datasets. From the total data, 80% for training, 10% for validation, and 10% for testing are used in the proposed system. The experimental results obtained accuracy, correctness, completeness, specificity, f1-score, and MCC of 98.26%, 98.26%, 98.26%, 99.13%, 98.25%, and 97.39% individually in overall.

In another study, Wong and Wang [123] developed a coronavirus detection mechanism from chest x-ray data called COVID-Net. The system generated a dataset COVIDx by combining and modifying two open-access datasets. In this study, the dataset consisted of a total of 13, 800 chest X-ray samples from 13,645 patients. The system considered three classes by combining bacterial and viral classes into a negative case. Among the total data, 90% was used for training and the rest 10% was utilized for validation. The proposed network obtained 92.4% accuracy in 10 iterations for test cases, and the sensitivity and precision of 80% and 88.9% were achieved in the case of COVID-19 class. Further, Khan et al. [124] proposed a deep CNN architecture named CoroNet for the diagnosis of COVID-19 infected patients from chest X-ray radiographs. In this system, a total of 1300 images were considered where 290 samples of COVID-19, 660 of bacterial pneumonia, 931 of viral pneumonia, and 1203 of normal patients. The dataset was split at a proportion of 80% and 20% for training and validation set respectively. The proposed system obtained accuracy, precision, sensitivity, and F1-score of 89.5%, 97%, 100%, and 98% respectively for COVID-19 class.

Recently, Rahimzadeh and Attar [125] proposed a modified CNN network for the diagnosis of novel coronavirus cases using X-ray samples. The system concatenated two well-known architecture of CNN named Xception and ResNet50V2 that make the system robust using multiple features extraction capability. Among the 15085 images, 180 were confirmed COVID-19, 6054 were pneumonia, and 8851 were normal cases. The scheme used 5-fold cross-validation for data partitioning. The network obtained accuracy of 99.50%, sensitivity of 80.53%, specificity of 99.56%, and precision of 35.27% for COVID-19 detection. Furthermore, Mukherjee et al. [126] proposed a system for the detection of novel coronavirus using shallow CNN in chest X-ray radiographs. The developed network is comparatively light-weight due to a small number of parameters. In this system, 130 positive COVID-19 cases, and 130 non-COVID cases were considered where the non-COVID cases include MERS, SARS, pneumonia, and normal chest X-rays. To obtain better performance, the dataset was split using 5-fold cross-validation. The performance of the system was evaluated by tuning the batch size of the CNN architecture. From the experimental results, it is found that the system obtained the highest accuracy, sensitivity, specificity, precision, F1-score, and AUC of 96.92%, 94.20%, 100%, 100%, 97.01%, and 99.22% respectively for batch size 50.

In another study, Li et al. [127] introduced a robust technique for automatic COVID-19 screening using discriminative cost-sensitive learning (DCSL). DCSL is formed with the combination of fine-grained classification and cost-sensitive learning. The used dataset consisted of 2,239 chest X-ray samples where 239 samples of COVID-19 cases, 1,000 samples from bacterial or viral pneumonia cases, and 1,000 samples of normal people. To obtain better performance, the dataset was partitioned using 5-fold cross-validation method. The proposed system achieved accuracy of 97.01%, precision of 97%, sensitivity of 97.09%, and F1-score of 96.98%. Khobahi et al. [128] developed a semi-supervised deep learning system based on Auto-Encoders named CoroNet to detect COVID-19 infected patients. The proposed system merged three open-access datasets for experiments. In this scheme, 18,529 images of different categories were used. Among the images, 99 samples were of COVID-19 classes, 9579 were of non-COVID pneumonia, and 8851 samples were related to healthy cases. The dataset was split in a proportion of 90% and 10% for training and testing set respectively. Overall, the accuracy, precision, recall, and F1-score of 93.50%, 93.63%, 93.50%, and 93.51% were achieved from the experiment. Moreover, Luz et al. [130] presented an efficient deep learning scheme named EfficientNet for the detection of coronavirus pattern from X-ray radiographs. The main advantage of EfficientNet is that it used fewer parameters, approximately 30 times less parameters than the pre-trained model. The system considered 30,663 images for experiment where 183 cases considered as COVID-19,

TABLE IV
SUMMARY OF DEEP LEARNING BASED COVID-19 DIAGNOSIS IN X-RAY IMAGES USING CUSTOMIZED NETWORK

| Authors | Data Sources | No. of images | No. of classes | Partitioning | Techniques | Performances (%) |
|---|---|---|---|---|---|---|
| Ozturk et al. [121] | COVID-19 X-ray image database [82], ChestX-ray8 [96] | 1127 (COVID=127, no-finding=500, pneumonia=500) | 3 (COVID, no-finding, pneumonia) | 5- fold cross-validation | DarkNet | Accuracy=98.08, Sensitivity=95.13, Specificity=95.3, Precision=98.03, F1-Score=96.51 |
| Ucar and Korkmaz [122] | COVID-19 X-ray image database [82], COVIDx Dataset [123], Kaggle chest X-ray pneumonia dataset [129] | 2839 (COVID-19=45, normal=1203, pneumonia=1591) | 3 (COVID-19, normal, pneumonia) | Training=80%, Testing=10%, Validation=10% | Bayes-SqueezeNet | Accuracy=98.26, Specificity=99.13, F1-Score=98.25, MCC=97.39, Correctness=98.26, Completeness=9826 |
| Wang and Wong [123] | COVID-19 X-ray image database [82], RSNA Pneumonia Detection Challenge dataset [90] | 13, 800 | 3 (COVID-19, non-COVID-19, normal) | Training=90%, Testing=10% | COVID-Net (CNN) | Accuracy= 92.4, Sensitivity= 80, Precision=88.9 |
| Khan et al. [124] | COVID-19 X-ray image database [82], Kaggle chest x-ray repository [87] | 1251 (COVID-19=284, normal=310, pneumonia bacterial=330, pneumonia viral= 327) | 4 (COVID-19, normal, pneumonia bacterial, pneumonia viral) | Training=80%, Validation=20% | CoroNet (CNN) | Accuracy=89.5, Sensitivity=100, Precision=97, F1-Score= 98 |
| Rahimzadeh and Attar [125] | COVID-19 X-ray image database [82], RSNA Pneumonia Detection Challenge dataset [90] | 15085 (COVID-19=180, pneumonia= 6054, normal= 8851) | 3 (COVID-19, pneumonia, normal) | 5- fold cross-validation | Concatenated CNN | Accuracy=99.50, Sensitivity=80.53, Specificity=99.56, Precision=35.27 |
| Mukherjee et al. [126] | covid-chestxray-dataset [82], Kaggle chest x-ray repository [87] | 260 (COVID-19=130, non-COVID=130) | 2 (COVID-19, non-COVID) | 5- fold cross-validation | Shallow CNN | Accuracy= 96.92, Sensitivity= 94.20, Specificity=100, Precision=100, F1-Score=97.01, AUC=99.22 |
| Li et al. [127] | COVID-19 X-ray image database [82], Kaggle dataset [83], Kermany et al. [84] | 2239 (COVID-19=239, pneumonia=1000, normal=1000) | 3 (COVID-19, pneumonia, normal) | 5-fold cross-validation | DCSL | Accuracy=97.01, Sensitivity=97.09, Precision=97, F1-Score=96.98 |
| Khobahi et al. [128] | COVID-19 X-ray image database [82], RSNA Pneumonia Detection Challenge dataset [90], COVIDx Dataset [123] | 18,529 (COVID-19=99, non-COVID-pneumonia=9579, healthy=8851) | 3 (COVID-19, non-COVID pneumonia, healthy) | Training=90%, Testing=10% | CoroNet (AutoEncoders) | Accuracy=93.50, Sensitivity=93.50, Precision=93.63, F1-Score=93.51 |
| Luz et al. [130] | COVID-19 X-ray image database [82], RSNA Pneumonia Detection Challenge dataset [90], COVIDx Dataset [123] | 30,663 (COVID-19=183, non-COVID pneumonia=14,348, healthy=16,132) | 3 (COVID-19, non-COVID pneumonia, healthy) | Training=90%, Testing=10% | EfficientNet | Accuracy=93.9, Sensitivity=96.8, Precision=100 |

| Alqudah et al. [131] | COVID-19 X-ray image database [82] | 71 (COVID-19=48, non-COVID-19=23) | 2 (COVID-19, non-COVID-19) | Training=70%, Testing=30% | CNN, SVM, RF | Accuracy=95.2, Sensitivity=93.3, Specificity=100, Precision= 100 |
|---|---|---|---|---|---|---|
| Farooq and Hafeez [132] | COVIDx Dataset [123] | 13, 800 | 4 (COVID-19, normal, bacterial, viral) | Training=90%, Testing=10% | COVID-ResNet (CNN) | Accuracy= 96.23, Sensitivity=100, Precision=100, F1-Score=100 |
| Afshar et al. [133] | COVIDx Dataset [123] | 13, 800 | 3 (COVID-19, normal, non-COVID-19) | Training=90%, Testing=10% | COVID-CAPS (Capsule Network) | Accuracy= 95.7, Sensitivity=90, Precision=95.8, AUC=97 |

16,132 images as normal cases, and 14,348 images as other pneumonia cases. The system obtained overall accuracy of 93.9%, sensitivity of 96.8%, and precision of 100%.

*2) Diagnosis Based on Single Source Data*

Alqudah et al. [131] suggested a hybrid method for the diagnosis of patients affected with coronavirus from X-ray data. The proposed system combined deep learning (CNN) and machine learning (SVM, RF) architecture. Deep learning was utilized both for feature extraction and classification task where machine learning was only used for classification task. In the system, 71 X-ray images of the chest were used in total where 48 are used for positive and 23 for negative. The system used 70% for training and 30% for testing and experiments were performed on several combinations like CNN-Softmax, CNN-SVM, and CNN-RF. The system obtained accuracy, sensitivity, specificity, and precision of 95.2%, 93.3%, 100%, and 100% respectively for CNN-Softmax classifier. Afterward, Farooq and Hafeez [132] presented a deep learning scheme with a pre-trained ResNet-50 network to detect the COVID-19 infected patients named COVID-ResNet. The proposed system implemented a residual neural network with 50 layers in total. In this study, the dataset comprised of 13,800 chest X-ray samples from 13,645 patients in total. The system achieved accuracy of 96.23% overall with 41 iterations using the COVIDx dataset. The other evaluation metrics like sensitivity, precision, and F1-score of 100%, 100%, and 100% are obtained considering the COVID-19 case only. Further, Afshar et al. [133] developed a capsule network based system named COVID-CAPS for the diagnosis of COVID-19 patients using X-ray samples. The proposed framework is well suited to work with a small dataset. The dataset used in the network was the COVIDx dataset, which is popular for COVID-19 research. The developed scheme used 13,800 chest x-ray images from 13,645 patients for experiments. Although there are four classes in the dataset, the system classifies the images into three classes considering bacterial and viral into one negative class. COVID-CAPS obtained accuracy, sensitivity, specificity, and AUC of 95.7%, 90%, 95.8%, and 0.97 respectively.

Table IV summarizes the aforementioned deep learning based COVID-19 diagnosis systems from X-ray samples using custom deep learning techniques and demonstrates some of the important factors, such as data sources, number of images and classes, data partitioning technique, diagnosis techniques, and the evaluation metrics of the developed systems.

## V. OPEN DISCUSSIONS, CHALLENGES AND FUTURE TRENDS

This section demonstrates the discussions of reviewed systems, challenges, and the possible future trends of deep learning based systems for COVID-19 diagnosis.

### A. Open Discussions

In this paper, 45 systems were reviewed where 23 systems used pre-trained model as a deep learning architecture, and the remaining 22 utilized custom deep learning framework. The results of the individual system are presented for explanation. Two popular imaging techniques CT and X-ray are used for data samples. Among the reviewed system, 25 systems are developed based on X-ray data and the rest 20 used CT samples. The majority of the systems used multiple source data and a few of them used single source data. We summarized the developed systems considering some features like the data sources, the number of images and classes, the data partitioning techniques, the used deep learning technique for diagnosis, and finally the evaluation metrics for performance measure. The data sources are the benchmark dataset or real-time data from hospital environment. Some of the systems used a huge number of images but the number of samples for COVID-19 cases is comparatively small. Both the binary and multi-class are considered throughout the review. With respect to data partitioning, some of the systems used cross-validation techniques, and others used hold-out method. Both the pre-trained model and custom deep learning architecture are taken into consideration. Almost all the systems used CNN or variants of CNN for diagnosis. Some common evaluation metrics like accuracy, sensitivity, specificity, precision, F1-score, AUC, etc. are utilized through the whole review.

The summary for the CT scan based COVID-19 diagnosis utilizing pre-trained model as well as customized deep learning technique is illustrated in Table I and Table III. It is evident from the results that most of the developed systems used real-time data from hospital environment of China and a few of the systems [61], [64], [103], [104], [109], [110], [111], and [119] used benchmark data. A few of the developed systems [66], [67], [68], [111], and [119] used data from a single source and the majority of the developed schemes used multiple source data. The datasets which are used multiple times in the reviewed systems are COVID-CT [112], and COVID-19 [113]. The reviewed systems which used maximum and minimum number of images for the experiment are [64], and [119] where the COVID-19 cases are of 32,230 (augmented data), and 51 respectively. In case of the number of classes to be classified, most of the developed systems considered binary class (COVID-19, and non-COVID-19) while some of them [60], [63], [64], [107], and [109] considered multiple classes (COVID-19, pneumonia, and normal). The 10-fold [119] was taken into consideration in some cases, whereas some developed systems [61], [62], [67], and [105] used random partitioning, and the majority of them considered hold-out method for data splitting. As far as performance is concerned, the system developed in [66] obtained 100% sensitivity. Among the reviewed systems, most of the frameworks [61], [62], [63], [65], [66], [103], [106], [107], [109] achieved comparatively higher accuracy, sensitivity, specificity, precision, F1-score and AUC having these measure (where applicable) greater than 90%. The highest accuracy of 99.51% and 99.68% were found at [66], [119] using pre-trained model and customized network respectively.

Table II and Table IV depict the X-ray based diagnosis of COVID-19 using pre-trained model with deep transfer learning and customized deep learning architecture. Our analysis revealed that almost all the developed systems used a common dataset COVID-19 X-ray image database [82]. Some systems used Kermany et al. [84] dataset, and COVIDx Dataset [123] frequently for diagnosis. All the proposed schemes utilized benchmark data for the experiment, no system applied real-time data. The systems introduced in [97], [131], [132], and [133] used single source data whereas the rest of the reviewed systems considered data from multiple sources. The framework demonstrated in [95] considered the highest number of images where the lowest case was used in [97]. The number of COVID-19 cases for [95], and [97] is 144, and 25 respectively. Although some systems considered the maximum number of images but the total number of images for COVID-19 is comparatively small. In terms of number of class consideration, some of the systems [70], [76], [124], and [131] considered 4 classes, a few of them [69], [75], [79], [81], [91], [94], etc. used 3 classes, and the remaining reviewed systems utilized binary class for experiment. The cross-validation technique such as 10-fold [69], [95], and 5-fold [80], [94], [121], [125], [126], [127] are used in some cases, and other systems considered hold-out method for data partitioning. In case of performance measure of the developed systems, 100% accuracy was achieved by the proposed systems in [70], 100% sensitivity in [70], [78], [97], [124], and [132], 100% specificity in [80], [97], [126], and [131], 100% precision in [70], [80], [97], [126], [130], [131], and [132], 100% F1-score [70], and [132], 100% AUC obtained in [80]. Most of the developed systems performed better in case of precision measurement, and the second best is obtained in case of sensitivity. Though the reviewed systems achieved comparatively better results for X-ray case both for pre-trained and custom network, the developed systems are not real-time tested with target people.

In comparing the pre-trained model with the custom network, some of the reviewed systems performed better for custom network. The performance of the developed systems varied depending on the dataset. The performance of the reviewed systems is not comparable as almost all the systems used different data sizes for the experiment. In terms of imaging modalities comparison, X-ray based systems performed comparatively better than CT based systems. But most of the X-ray based frameworks used benchmark data while the real-time data from hospital environment is considered in case of CT based systems. It is envisaged that the systems introduced using CT samples are applicable for real-time testing but the X-ray based proposed schemes need real-time testing with target people before application

### B. Challenges and Future Trends

There are many unique challenges for applying deep learning techniques and algorithms for the detection of novel coronavirus (COVID-19).

While deep learning techniques are highly automatable, it needs a large set of data to develop a robust system for diagnosis purpose. As the COVID-19 is very new to research, the lack of standard data is a major challenge for diagnosis. On the other hand, the available imaging data for COVID-19 patients are incomplete, noisy, ambiguous, and inaccurate labels in some cases. To train a deep learning architecture with such massive and diverse data sets is very complex and there is a need to resolve a variety of problems like data redundancy, sparsity as well as missing values. Almost all the reviewed systems used different data sets for the experiment. The developed systems collected data from internet sources, prepared the data in their way, and finally evaluated their systems using evaluation metrics. For this reason, it is quite difficult to conclude definitively which system yields the best result for COVID-19 detection.

The further challenges for the COVID-19 detection systems are the imbalance in the dataset samples. It is a critical issue as there are a few COVID-19 samples both CT and X-ray whereas pneumonia and normal cases contain a huge number of samples than COVID-19 cases. It is evident from the reviewed systems proposed in [61], [62], [78], [95], etc. The imbalance in data very often raises bias during the training phase of deep learning technique. With the fewer number of positive samples, it has become increasingly difficult to balance out the target sample. The lack of confidence interval is another challenge found in deep learning based COVID-19 diagnosis systems. Deep learning architecture provides the output as prediction confidence whereas the output indicator of a particular neuron is considered as a single probability. For COVID-19 diagnosis, the lack of confidence interval across a predicted value is usually not desirable.

To overcome these challenges, researches may consider the design of optimized deep learning algorithms that can easily cope up with a small number of data [26]. A shallow long short-term memory (LSTM) [134] is utilized to solve the limitations of a small dataset. In the absence of large-scale training datasets, leveraging the current deep learning architecture as feature extractors and then conducting more learning on those attributes using end to end manner [135], [136] is a more encouraging path. Freezing is a technique that provides the facilities to shrink the number of parameters in deep learning architecture where the reduced parameters are hired from another network trained for similar purposes. While the number of parameters would reduce, it might be possible to achieve good performance from a small number of COVID-19 cases [137], [138]. Ensemble learning [139], [140], and multi-task learning [110], [141] are better suited for COVID-19 diagnosis in the context of a small number of data. In case of ensemble learning, multiple architectures are developed instead of a single network and finally, the results of each network are combined. In multi-task architecture, diverse tasks are combined to take the facility of data annotations from one another.

Additionally, synthetic data generation might be a possible solution to overcome the challenges of deep learning based COVID-19 detection systems. The most used techniques for data generation are data augmentation and GANs which are frequently utilized to solve the class imbalance problem. The data augmentation technique [142], [143] generate new lesions from the given COVID-19 samples using flipping, rotation, cropping, random noise addition, etc. from the given images. But the overfitting problem may arise in case of augmented data. GANs are the most sophisticated techniques for realistic synthetic data generation. From a small number of COVID-19 samples, GANs [70], [144] generate a large number of images that are used to train a deep learning system for the novel virus diagnosis. Further, weakly supervised deep learning methods [145], [146] would be a probable solution for limited training data. Furthermore, as the manually labeling of COVID-19 imaging data is costly, and lengthy process, the use of self-supervised deep learning techniques [147], [148] are highly recommended.

## VI. CONCLUSION

The COVID-19 is still an ongoing pandemic that is creating new records in terms of cumulative and daily numbers for global infection and death. Deep learning based automatic diagnosis of COVID-19 which provides consistent and accurate solutions, has played a significant role to assist with the diagnosis of this disease. This paper presents the recent works for COVID-19 diagnosis purposes using deep learning techniques from two types of imaging techniques like CT and X-ray samples. The review describes the systems which are developed based on pre-trained model with deep transfer learning and customized deep learning architecture for COVID-19 diagnosis. Two-leveled taxonomy was presented which explores the perspectives of deep learning techniques and imaging modalities. This paper outlines all the sources of used datasets which can be easily understood and accessed by the research community. The major challenge of the COVID-19 diagnosis systems based on deep learning is the lack of gold standard. Furthermore, the possible solutions to overcome the current challenges are recommended that might encourage the researchers who would like to contribute in this area. It is prudent to note that the deep learning techniques with imaging modalities offer only partial details about the infected patients. However, it is not really envisaged in the present state that the role of physicians or clinicians in clinical diagnosis can be replaced by deep learning techniques. In the near future, it is hoped that deep learning experts with radiologists would provide appropriate support systems for identifying the COVID-19 infected patients.